\documentclass[twocolumn,secnumarabic,amssymb, nobibnotes, aps, pra]{revtex4-2}
\usepackage{color}

\newcommand{\macro}[1]{\texttt{\textbackslash#1}}
\newcommand{\m}[1]{\macro{#1}}

\newcommand{\vect}[1]{{\mbox{\boldmath $#1$}}}
\newcommand{\red}[1]{\textcolor{black}{#1}}

\usepackage{amsmath,amssymb}
\usepackage{bm}
\usepackage{graphicx}
\begin{document}

\title{Cavity magnomechanical coupling with coupled magnon modes \\in a synthetic antiferromagnet}

\author{Motoki Asano$^1$}
\author{Hiroki Matsumoto$^2$}
\author{Masamitsu Hayashi$^{2,3}$}
\author{Daiki Hatanaka$^1$}

\begin{abstract}
On-chip cavity magnomechanics is an emerging field exploring acoustic and magnonic functionalities of various ferromagnetic materials and structures using strongly confined phonons. It is expected that such cavity magnomechanics can be extended to multilayer ferromagnets, especially synthetic antiferromagnets (SAFs) that exhibit zero net magnetization through interlayer exchange coupling. However, the conventional theoretical framework for a single ferromagnet cannot be used directly because of the antiferromagnetic magnetization dynamics associated with the interlayer exchange coupling. In this paper, we theoretically investigate phonon-magnon coupling with a three-layer SAF. Our formulation of the phonon-magnon coupling constants reveals that the acoustic (optical) magnon mode dominantly couples to the cavity phonon when the magnetization angles in the two ferromagnetic layers are antiparallel (\red{near-}orthogonal). Moreover, numerical calculations including the effects of dipole-dipole interactions and in-plane uniaxial magnetic anisotropy allow us to predict phonon frequency shifts and linewidth broadening that can be detected in experiments. These theoretical insights would greatly help us to make a strategy for bringing the system into the strong coupling regime and to devise novel control protocols in analogy to cavity quantum electrodynamics and cavity optomechanics.
\end{abstract}

\address{$^1$NTT Basic Research Laboratories, NTT Corporation, Atsugi, Kanagawa 243-0198, Japan\\
$^2$Department of Physics, The University of Tokyo, Hongo, Tokyo 113-0033, Japan\\
$^3$Trans-scale quantum science institute (TSQS), The University of Tokyo, Hongo, Tokyo 113-0033, Japan
}
\date{\today}
\maketitle
\section{Introduction}
The ability to manipulate interactions between acoustic waves ({\it i.e.,} phonon) and spin precession in ferromagnetic materials ({\it i.e.,} magnon) has attracted significant interest in magnomechanics \cite{weiler2011elastically,dreher2012surface,thevenard2014surface,kikkawa2016magnon,kobayashi2017spin,labanowski2018voltage,sasaki2019surface,an2020coherent,hernandez2020large,shah2020giant,xu2020nonreciprocal,sasaki2021magnetization}. In particular, cavity magnomechanics is an emerging field of study in which phonons are strongly confined in a cavity to significantly enhance the phonon-magnon coupling. This cavity enhancement offers functionalities such as coherent energy conversion from phonons to magnons and {\it vice versa} \cite{zhang2016cavity,hatanaka2022chip,potts2023dynamical}, control of dynamical back-action \cite{potts2021dynamical}, back-action evading measurement \cite{potts2023dynamical}, and mechanical bistability \cite{shen2022mechanical}. The initial studies used a yttrium iron garnet (YIG) sphere where a radially breathing phonon mode and a collective excitation of magnetizations referred to as the Kittel mode interact with each other \cite{zhang2016cavity,potts2023dynamical}. The cavity magnomechanical coupling in the YIG sphere is regarded as a dispersive interaction in which the energy of the magnon modes is modulated with respect to the strain field associated with the cavity phonon mode. This theoretical description helps us not only to understand certain experimental results but also to devise a novel control protocols \cite{li2018magnon,potts2020magnon,lu2021exceptional,sarma2021cavity} in analogy to those of cavity quantum electrodynamics (QED) \cite{walther2006cavity} and cavity optomechanics \cite{aspelmeyer2014cavity}.

On the other hand, the recent progress of on-chip phononic devices has expanded the cavity magnomechanics framework to a variety of materials and structures \cite{hwang2020enhancement,hatanaka2022chip,hatanaka2023phononic}. This framework allows a resonant magnomechanical coupling where the phonon resonance frequency matches the magnon resonance frequency and provides tailorable phononic and magnonic functionalities. Recently, a phononic crystal cavity was developed that strongly confines the strain field inside a wavelength-scale cavity and induces strain fields along multiple axes \cite{hatanaka2023phononic}.  What comes next in cavity magnomechanics is functionalizing magnon modes by engineering magnetic structures in the phonon cavity similarly to magnomechanics with propagating phonons \cite{kuss2021nonreciprocal,matsumoto2022large}. Here, a synthetic antiferromagnet (SAF), which consists of a series of ferromagnetic layers and non-magnetic spacers deposited on top of one another, is a strong candidate for on-chip cavity magnomechanics, because the interlayer exchange coupling induces unique magnetic properties, such as an anti-parallel ({\it i.e.,} antiferromagnetic) configuration of magnetizations and coupled magnon modes \cite{parkin1990oscillations}. The antiferromagnetic magnetization configuration among the ferromagnetic layers and magnon modes in SAFs can be manipulated even when the net magnetization is zero, which is of great importance in the field of spintronics \cite{duine2018synthetic}. In SAFs, the magnetization precessions in the ferromagnetic layers are coupled, resulting in in-phase and anti-phase ferromagnetic resonance precessions referred to as the acoustic and optical magnon mode, respectively. \red{Thus, cavity magnomechanics with a SAF allows the introduction of multiple magnon modes, which can further expand the control protocol in the same way as the use of multiple optical modes does in cavity optomechanics \cite{lee2015multimode,nielsen2017multimode}. Moreover, the SAF structure shows a rich mode tunability where the resonance frequency and the magnetoelastic coupling constant can be controlled via an external magnetic field.} However, the theoretical framework for cavity magnomechanics that was developed in the previous studies \cite{hatanaka2022chip,hatanaka2023phononic} is not directly applicable to SAFs because the magnon mode may couple to cavity phonons in a different way than the Kittel mode in a single ferromagnet does.

In this paper, we theoretically investigate the phonon-magnon coupling in a phonon cavity and a three-layer SAF as a model system. The three-layer SAF consists of two ferromagnetic layers with the same ferromagnetic material sandwiching a non-magnetic spacer. As a typical property, the two magnetization vectors in the ferromagnets are aligned anti-parallel ({\it i.e.,} in the antiferromagnetic configuration). First, we provide a general description for cavity magnomechanics with a three-layer SAF coupled to an arbitrary phonon field via a magnetoelastic coupling in Sec II. Then, in Sec III, we apply this formalism to a simplified model, {\it i.e.,} an SAF integrated on a surface acoustic wave (SAW) resonator. The phonon-magnon coupling constants for the acoustic and optical magnon modes are formulated to clarify the dependence of these couplings on the angle of the external magnetic field and the relative angle between the magnetization directions. Moreover, the effects of the dipole-dipole interaction and in-plane uniaxial magnetic anisotropy on the magnon modes are numerically investigated. Finally, we present a synthesis to obtain the phonon-magnon coupling constant, illustrate the difference between phonon-magnon and photon-magnon couplings, and discuss prospects for reaching the strong coupling regime in Sec. IV.

\section{General Formulation}
\subsection{Setup}
Figure 1(a) shows a schematic illustration of a cavity magnomechanical system where the cavity phonon field interacts with the SAF. This model enables us to discuss cavity magnomechanics by considering the simplest SAF structure consisting of two identical ferromagnetic layers (having the same saturation magnetization and the same thickness) separated by a non-magnetic spacer. The magnetization dynamics in the ferromagnetic layers can be described by using macrospins, which provides a coarse-grained physical interpretation. The equilibrium directions of the magnetizations are anti-parallel because of the antiferromagnetic interlayer exchange coupling. In addition, the interlayer exchange coupling causes the two magnetizations to show in-phase and anti-phase precessions, which respectively correspond to acoustic and optical magnons when the interlayer exchange interaction dominantly determines the magnetization dynamics in the three-layer SAF. Moreover, the self and interlayer dipole-dipole interactions can induce mode hybridization between the acoustic and optical magnons, as will be discussed in Sec III \cite{shiota2020tunable,sud2020tunable}.

For simplicity, we impose three assumptions on our cavity magnomechanical model by referring to the previous experimental configuration \cite{hatanaka2022chip}: (i) the coupling between the cavity phonon mode and the magnon mode originates from the magnetoelastic coupling, where we ignore other contributions such as magneto-rotation, spin-rotation and spin-vorticity coupling \cite{xu2020nonreciprocal,kuss2021symmetry}; (ii) the ferromagnetic film is thin ($<$10 nm) with no perpendicular magnetic anisotropy, whereby the equilibrium directions of the magnetizations are in the film plane because of the strong shape anisotropy normal to the film; (iii) the total thickness of the SAF structure is much smaller than the acoustic wavelength, which results in the equivalent magnetoelastic couplings to the macrospins because the strain wave uniformly penetrates the ferromagnetic layers. Figure 1(b) shows a diagram of the cavity magnomechanical system. 

To describe the dynamics of the magnon modes, a set of coordinates referred to as the ``spin frame" and denoted by $(x_{i},y_i,z_i)$ $(i=1,2)$ is introduced [see Fig. 1(c)]. The normalized magnetization vector of the $i$th layer is given by $\vect{m}_i=(m_{ix},m_{iy},m_{iz})$, where the $z$ direction corresponds to the magnetization direction in equilibrium with an angle $\varphi_i$ from the $X$ axis. Note that this magnetization vector is normalized by the saturation magnetization $M_\mathrm{S}$ so that $|\vect{m}_i|=1$. By neglecting the terms of order $\mathcal{O}(m_{is}^2)$ $(s=x, y)$ and assuming that $m_{iz}\approx 1$ and $m_{ix}, m_{iy} \ll 1$, we can linearize the magnetization dynamics, which is referred to as the linear precession approximation. Although the interlayer exchange coupling makes the magnetization directions in equilibrium anti-parallel (antiferromagnetic: $|\varphi_1-\varphi_2|=\pi$), increasing the in-plane magnetic field aligns them so that they become parallel (ferromagnetic: $|\varphi_1-\varphi_2|=0$). Thus, the relative angles $\varphi_1-\varphi_2$ can be adjusted by tuning the strength of the in-plane external magnetic field.

\begin{figure}[h]
  \centering
  \includegraphics[width=8cm]{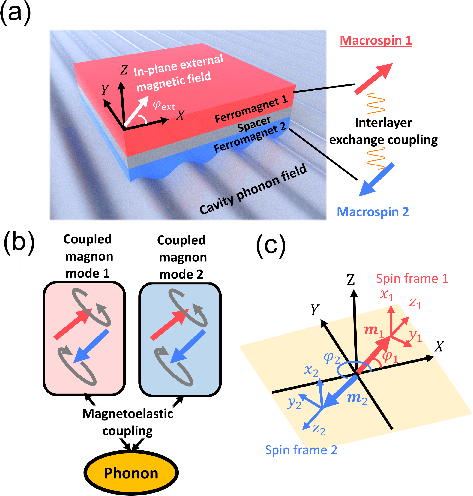}
  \caption{(a) Conceptual illustration of setup with SAF and cavity phonon field. (b) Diagram of mode coupling between phonons and coupled magnon modes. (c) Macrospin vectors in the laboratory frame $(X, Y, Z)$ and spin frames $(x_i, y_i, z_i)$. }
\end{figure}

\subsection{Dynamics of magnon modes}
The dynamics of macrospins in each ferromagnetic layer is described by the Landau-Lifshitz-Gilbert  (LLG) equation,
\begin{align}
\dot{\vect{m}}_i=-\gamma \vect{m}_i\times \mu_0\vect{H}_{\mathrm{eff},i}+\alpha \vect{m}_i\times\dot{\vect{m}}_i, \label{LLG1}
\end{align}
where $\gamma$, $\mu_0$, and $\alpha$ are the gyromagnetic ratio, the magnetic permeability of the vacuum, and the Gilbert damping coefficient, respectively. The effective magnetic field, $\mu_0\vect{H}_{\mathrm{eff},i}$, is given by $\mu_0\vect{H}_{\mathrm{eff},i}=- M_\mathrm{S}^{-1}\nabla_{m_i}E$, where $E$ is the magnetic free energy density shown in Appendix A. This effective magnetic field can be decomposed into
\begin{align}
\mu_0\vect{H}_{\mathrm{eff},i}=&\mu_0\vect{H}_{\mathrm{ext},i}+\mu_0\vect{H}_{\mathrm{dem},i}+\mu_0\vect{H}_{\mathrm{int},i}+\mu_0\vect{H}_{\mathrm{dip},i}\nonumber\\
&+\mu_0\vect{H}_{\mathrm{uni},i}+\mu_0\vect{H}_{\mathrm{PM},i}, \label{effH}
\end{align}
where $\mu_0\vect{H}_{\mathrm{ext},i}$ is the in-plane external magnetic field, $\mu_0\vect{H}_{\mathrm{dem},i}$ is the demagnetizing field, $\mu_0\vect{H}_{\mathrm{int},i}$ is the interlayer exchange field, $\mu_0\vect{H}_\mathrm{dip}$ is the sum of the self and interlayer dipolar fields, $\mu_0\vect{H}_{\mathrm{uni},i}$ is the in-plane uniaxial magnetic anisotropy field, and $\mu_0\vect{H}_{\mathrm{PM},i}$ is the effective field from the magnetoelastic coupling, respectively. The detailed expressions are given in Appendix A. 

By applying the linear precession approximation ({\it i.e.,} $m_{iz}\approx 1$ and $m_{ix}, m_{iy} \ll 1$), the LLG equation (\ref{LLG1}) can be linearized as
\begin{align}
\dot{\vect{m}}=A_\mathrm{M}\vect{m}+\vect{b}_\mathrm{M}, \label{LLG2}
\end{align}
where $\vect{m}$ denotes a four-dimensional vector $(m_{1x},m_{1y},m_{2x},m_{2y})$ for describing spin precession in both ferromagnets, and $A_\mathrm{M}$ and $\vect{b}_\mathrm{M}$ are respectively a $4\times4$ matrix and a four-dimensional vector. The eigenenergy, $\epsilon_\mu$, and eigenvectors of $A_\mathrm{M}$,  $\vect{v}_\mu$, are calculated by diagonalizing $A_\mathrm{M}$ with $P^{-1}A_\mathrm{M}P=\mathrm{diag}[\epsilon_1,\epsilon_2,\epsilon_3,\epsilon_4]$ with $P\equiv (\vect{v}_1,\vect{v}_2,\vect{v}_3,\vect{v}_4)$. Thus, Eq. (\ref{LLG2}) transforms into
\begin{align}
\dot{\tilde{\vect{m}}}=\mathrm{diag}[\epsilon_1,\epsilon_2,\epsilon_3,\epsilon_4]\tilde{\vect{m}}+\tilde{\vect{b}}_\mathrm{M},
\end{align}
where $\tilde{\vect{m}}\equiv P^{-1}\vect{m}$ with the vectors of the magnon modes, $\tilde{\vect{m}}=(\tilde{m}_1,\tilde{m}_2,\tilde{m}_3,\tilde{m}_4)$, and $\tilde{\vect{b}}_\mathrm{M}=P^{-1}\vect{b}_\mathrm{M}$. For later use, we define $\omega_{\mathrm{M},\mu}=\mathrm{Im}[\epsilon_\mu]$ and $\kappa_{\mathrm{M},\mu}=\mathrm{Re}[\epsilon_\mu]/2$, which are the magnon resonance frequency and damping constant, respectively. 

To discuss the phonon-magnon coupling in the case of an arbitrary cavity strain field, $\vect{b}_\mathrm{M}$ can be expressed as the sum of the products of the strain $\varepsilon_{ij}$ and magnetoelastic coupling terms,
\begin{align}
\vect{b}_\mathrm{M}=\sum_{i,j=X,Y,Z}\varepsilon_{ij}\vect{q}_{ij}(\varphi_1,\varphi_2),
\end{align}
where $\vect{q}_{ij}(\varphi_1,\varphi_2)$ is a four-dimensional vector including the magnetoelastic coupling strength (see Appendix B). This decomposition based on $\vect{q}_{ij}$ is useful for examining the phonon-magnon coupling in an arbitrary cavity strain field. When the SAW resonator holds only a single longitudinal strain component $\varepsilon_{ss}$, we can simply express the term of phonon-magnon coupling by $\vect{b}_\mathrm{M} = \varepsilon_{ss} \vect{q}_{ss}$. In the more complicated case, the cavity field consists of other strain components also including shear strains and thus appropriate terms of $\vect{q}_{ij}$ ($i, j = X, Y, Z$).

We can decompose the displacement field $\vect{u}(\vect{r})$ into spatial components $\psi_s(\vect{r})$ and a temporal component $U(t)$ as $\vect{u}(\vect{r})=(\psi_X(\vect{r}),\psi_Y(\vect{r}),\psi_Z(\vect{r}))U(t)$. The strain field $\varepsilon_{sw} \sim (\partial_s u_w + \partial_w u_s)/2$ can then be described as $\varepsilon_{sw}=\Psi_{sw}(\vect{r})U(t)$, where $\Psi_{sw}(\vect{r})\equiv (\partial_s \psi_w+\partial_w \psi_s)/2$. In the same manner, the magnon modes can be decomposed into $\tilde{m}_\mu=\Phi_\mu(\vect{r})M_\mu(t)$. Thus, Eq. (\ref{LLG2}) can be simplified as,
\begin{align}
\dot{M}_\mu(t)=\epsilon_\mu M_\mu(t) +g_{\mathrm{MP},\mu}U(t), \label{LLG3}
\end{align}
where
\begin{align}
g_{\mathrm{MP},\mu}=\sum_{j,k=X,Y,Z}\frac{\int\mathrm{d}^3\vect{r}\Phi_\mu(\vect{r})\Psi_{jk}(\vect{r})}{\int\mathrm{d}^3\vect{r}\Phi^2_\mu(\vect{r})}\left[P^{-1}\vect{q}_{jk}\right]_\mu, \label{int1}
\end{align}
showing the strength of magnetoelastic coupling between the $\mu$th magnon mode and the cavity phonon mode.

\subsection{Dynamics of phonon modes}
The dynamics of the phonon modes is described by the equation of motion,
\begin{align}
\ddot{U}(t)+\kappa_\mathrm{P} \dot{U}(t)+\omega_\mathrm{P}^2 U(t)=\frac{\int\mathrm{d}^3\vect{r}\vect{\psi(\vect{r})}\cdot\vect{f}_\mathrm{PM}}{\rho\int\mathrm{d}^3\vect{r}|\vect{\psi}(\vect{r})|^2}, \label{EQM1}
\end{align}
where $\rho$, $\kappa_\mathrm{P}$ and $\omega_\mathrm{P}$ are the density of the host material making up the phonon cavity, the acoustic damping, and the resonance frequency in the cavity phonon mode, respectively, and $\vect{f}_\mathrm{PM}$ is the dynamic magnetostrictive force given by
\begin{align}
[\vect{f}_\mathrm{PM}]_\mu=&\sum_{j=X,Y,Z}\partial_j \partial_{\varepsilon_{\mu j}} E_\mathrm{PM},
\end{align}
where $E_\mathrm{PM}$ is the energy density from the magnetoelastic coupling. In the linear precession regime, the equation of motion, Eq. (\ref{EQM1}), can be rewritten as
\begin{align}
\ddot{U}(t)+\kappa_\mathrm{P} \dot{U}(t)+\omega_\mathrm{P}^2 U(t)=\sum_\mu g_{\mathrm{PM},\mu} M_\mu(t).\label{EQM2}
\end{align}
Here, the strength of the magnetoelastic coupling $g_{\mathrm{PM},\mu}$ is defined as
\begin{align}
\frac{\int\mathrm{d}^3\vect{r} \bar{\vect{q}}\cdot P\tilde{\vect{m}}}{\rho\int\mathrm{d}^3\vect{r}|\vect{\psi}(\vect{r})|^2}=\sum_\mu g_{\mathrm{PM},\mu} M_\mu(t), \label{int2}
\end{align}
using a four-dimensional vector operator $\bar{\vect{q}}$, which includes a spatial derivative characterizing how the strain field couples to the spatial derivative of the dynamic magnetization (the detailed expression is given in Appendix C).

\subsection{Magnomechanical coupling}
The coupled dynamics between the magnon modes in the SAF and the cavity phonon mode are determined by solving the linear coupled mode equations, Eqs. (\ref{LLG3}) and (\ref{EQM2}). Bringing them into the frequency domain, we obtain a linear algebraic equation, $B(U,M_1,M_2,M_3,M_4)^T=\vect{f}_d$, where $\vect{f}_d$ is the driving field vector. $B$ is a $5\times5$ matrix,
\begin{align}
&B=\left(\begin{array}{ccccc}
\chi_P&-g_{\mathrm{PM},1}&-g_{\mathrm{PM},2}&-g_{\mathrm{PM},3}&-g_{\mathrm{PM},4}\\
-g_{\mathrm{MP},1}&\chi_{\mathrm{M},1}&0&0&0\\
-g_{\mathrm{MP},2}&0&\chi_{\mathrm{M},2}&0&0\\
-g_{\mathrm{MP},3}&0&0&\chi_{\mathrm{M},3}&0\\
-g_{\mathrm{MP},4}&0&0&0&\chi_{\mathrm{M},4}\\
\end{array}\right),\end{align}
where $\chi_\mathrm{P}=-\omega^2-i\omega \kappa_\mathrm{P} +\omega_\mathrm{P}^2$ and $\chi_{\mathrm{M},\mu}=-i\omega -\epsilon_\mu$ are the susceptibilities in the phonon and magnon modes, respectively. In a typical setup of on-chip cavity magnomechanics, the frequency shift, $\delta\omega$, and the linewidth broadening, $\delta\kappa$, under acoustic driving, {\it i.e.,} $\vect{f}_d=(f_0,0,0,0,0)^T$, are measured experimentally \cite{hatanaka2022chip,hatanaka2023phononic}. Accordingly, the modulated phonon susceptibility $\tilde{\chi}_\mathrm{P}$ is given by
\begin{align}
\tilde{\chi}_\mathrm{P}=&{\chi}_\mathrm{P}-\sum_{\mu}\chi^{-1}_{\mathrm{M},\mu}g_{\mathrm{PM},\mu}g_{\mathrm{MP},\mu}\nonumber\\
=&-2\omega_\mathrm{P}\left[\omega-\omega_\mathrm{P}+\delta\omega+i\left(\frac{\kappa_\mathrm{P}+\delta\kappa}{2}\right)\right],
\end{align}
where 
\begin{align}
\delta\omega=&\sum_{\mu}\mathrm{Re}\left[\frac{g^2_{0,\mu}}{i\omega+\epsilon_\mu}\right], \label{freq}\\
\delta\kappa=&\frac{1}{2}\sum_{\mu}\mathrm{Im}\left[\frac{g^2_{0,\mu}}{i\omega+\epsilon_\mu}\right]. \label{linewidth}
\end{align}
Here, the magnomechanical coupling constant is given by a symmetrical form,
\begin{align}
g_{0,\mu}=\sqrt{\frac{g_{\mathrm{PM},\mu}g_{\mathrm{MP},\mu}}{2\omega_\mathrm{P}}}.
\end{align}
Note that $g_{0,\mu}$ is in units of angular frequency, and determines the energy transfer rates between the phonon and magnon modes. Thus, the strong coupling regime is where $g_{0,\mu}$ is larger than $\kappa_\mathrm{P}$ and $\kappa_\mathrm{M}$ (see Sec. IV).

When the angular frequency of the driving force is close to the cavity resonance, $\omega\approx \omega_\mathrm{P}$, the phonon frequency shift and the linewidth broadening are inversely proportional to the frequency difference between the phonon and magnon modes. When the $s$th magnon mode satisfies the resonance condition, {\it i.e.,} $\omega_{\mathrm{M},s}=-\omega_\mathrm{P}$ and the other magnon modes show $|\omega_{\mathrm{M},\mu}-\omega_\mathrm{P}|\gg 0$  $(\mu \neq s)$, the modified susceptibility simplifies to as $|\tilde{\chi}_\mathrm{P}|\approx \kappa_\mathrm{P}\omega_\mathrm{P}\sqrt{1+C^2}$, where $C
\equiv 4g^2_{0,s}/\kappa_\mathrm{P} \kappa_{M,s}$ denotes the cooperativity. In analogy to cavity QED and optomechanics \cite{walther2006cavity,aspelmeyer2014cavity}, $C\gg 1$ means that the externally driven phonon energy can be coherently converted. 

\section{A simple cavity model: SAW resonator}
\subsection{Setup}
Here, we investigate the phonon-magnon coupling of a simple phonon cavity model based on a SAW resonator. SAW resonators are often used to apply dynamic strain fields with a widely tunable phonon resonance frequency from several MHz to about 10 GHz, by designing the period of the grating in the inter digital transducers (IDTs) and Bragg reflectors (BRs) \cite{xu2018high,shao2019phononic}. Figure 2(a) shows a schematic diagram of the cavity magnomechanics setup with the SAF and SAW resonator. For simplicity, we assume that the longitudinal strain along the $X$ axis (SAW propagation axis) is dominant and the other strain components can be neglected.

\begin{figure}[htbp]
  \centering
  \includegraphics[width=8cm]{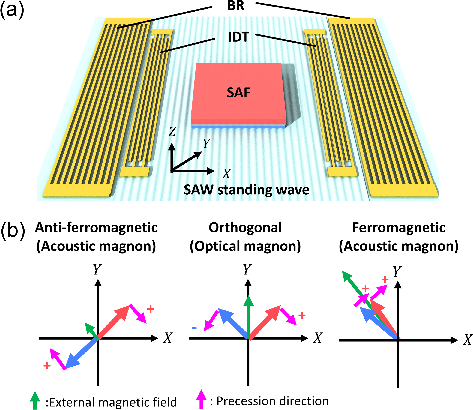}
  \caption{(a) Cavity magnomechanics setup with SAW resonator and SAF. The dynamic strain field in the phonon mode is excited by the inter digital transducer (IDT) and is confined between the two Bragg reflectors (BRs). (b) Schematic diagram of acoustic and optical magnon modes dominantly coupled to the strain field with different external magnetic fields.}
\end{figure}

\subsection{Analytical formulation of phonon-magnon coupling constant}
Here, let us formulate the phonon-magnon coupling constant in a three-layer SAF on a SAW resonator. For simplicity, we neglect the dipole-dipole interaction and the in-plane uniaxial magnetic anisotropy. In other words, we consider only $\mu_0 \vect{H}_{\mathrm{ext}, i}$, $\mu_0\vect{H}_{\mathrm{dem}, i}$, and $\mu_0\vect{H}_{\mathrm{int}, i}$ in Eq. (2). The magnetic energy density is expressed by
\begin{align}
E&\approx -\mu_0H_\mathrm{ext}M_\mathrm{S}\sum_{i=1,2}\left[\cos(\varphi_\mathrm{ext}-\varphi_i)+m_{iy}\sin(\varphi_\mathrm{ext}-\varphi_i)\right]\nonumber\\
&+\frac{\mu_0M_\mathrm{S}^2}{2}(m^2_{1x}+m^2_{2x})\nonumber\\
&+\mu_0H_\mathrm{E}M_\mathrm{S}[\cos(\varphi_1-\varphi_2)+\sum_{i=1,2}(-1)^i m_{iy}\sin(\varphi_1-\varphi_2)],
\end{align}
where $\mu_0 H_\mathrm{ext}$ and $\varphi_\mathrm{ext}$ are the strength and angle of the in-plane external field, respectively, and $\mu_0H_\mathrm{E}$ is the interlayer exchange magnetic field. Here, the equilibrium magnetization angles, $\varphi_1$ and $\varphi_2$, are determined so that $E(\varphi_1,\varphi_2)$ takes a minimum value. From the symmetry of the SAF structure, we assume $\varphi_1 = \varphi_\mathrm{ext} + \varphi_0$ and $\varphi_2 = \varphi_\mathrm{ext} - \varphi_0$, where $\varphi_0=\cos^{-1}\left(\frac{H_\mathrm{ext}}{2H_\mathrm{E}}\right)$. The magnon modes can be derived by diagonalizing $A_\mathrm{M}$:
\begin{align}
A_\mathrm{M}=D_\mathrm{M}\left(\begin{array}{cccc}
0&H_\mathrm{E}&0&a_c\\
-M_\mathrm{S}-H_\mathrm{E}&0&-H_\mathrm{E}&0\\
0&a_c&0&H_\mathrm{E}\\
-H_\mathrm{E}&0&-M_\mathrm{S}-H_\mathrm{E}&0
\end{array}\right), \label{SAW_AM}
\end{align}
where $a_c\equiv H_\mathrm{E}\cos2\varphi_0$, and $D_\mathrm{M}$ is given in Appendix B. Here, we emphasize that the eigenmodes of $A_\mathrm{M}$ can be written in an in-phase form $(m_{1x},m_{1y})=(m_{2x},m_{2y})$, {\it i.e.,} an acoustic magnon mode, and an anti-phase form $(m_{1x},m_{1y})=-(m_{2x},m_{2y})$, {\it i.e.,} an optical magnon mode (see Appendix D). To derive the phonon-magnon coupling constants, we assume that the phonon mode and coupled magnon modes are in the standing wave forms, $\psi_i(\vect{r})=F(Z)\sin kX$ ($i=X, Y, Z$) and $\Phi_\mu(\vect{r})=F(Z)\cos kX$ ($\mu$=1,2,3,4), where $k$ and $\lambda$ are the wavenumber and wavelength of the SAW, respectively, and $F(Z)$ is the spatial distribution of the phonon mode along the $Z$ direction. By integrating the distribution along Eqs. (\ref{int1}) and (\ref{int2}) and taking $\alpha \rightarrow 0$, we define the coupling constants as
\begin{align}
g_\mathrm{ac}=&k\left|b_1\sin2\varphi_\mathrm{ext}\red{\frac{\cos2\varphi_0}{\sqrt{|\cos\varphi_0|}}}\right|\red{\sqrt[4]{\frac{2H_\mathrm{E}+M_\mathrm{S}}{2H_\mathrm{E}}}}\sqrt{\frac{\gamma M_\mathrm{S}}{4\rho\omega_\mathrm{P}}\frac{L_\mathrm{M}}{L_\mathrm{A}}\frac{d}{\lambda}}\label{g_ac},\\
g_\mathrm{opt}=&k\left|b_1\cos2\varphi_\mathrm{ext}\red{\frac{\sin2\varphi_0}{\sqrt{||\sin\varphi_0}}}\right|\red{\sqrt[4]{\frac{M_\mathrm{S}}{H_\mathrm{E}}}}\sqrt{\frac{\gamma M_\mathrm{S}}{4\rho\omega_\mathrm{P}}\frac{L_\mathrm{M}}{L_\mathrm{A}}\frac{d}{\lambda}},\label{g_opt}
\end{align}
where $L_\mathrm{A}$ and $L_\mathrm{M}$ are the respective lengths of the SAW resonator and the ferromagnetic film along the $X$ direction, and $d$ is the total thickness of the SAF structure. From the angle dependence, we find that the acoustic magnon mode dominantly couples to the phonon mode when the relative magnetization angle, \red{$\varphi_1-\varphi_2\approx 0$ and 180 deg} at $\varphi_\mathrm{ext}=45(2l-1)$ deg ($l=1,2,3,4$) \red{[see Appendix D]}. Thus, both the antiferromagnetic and ferromagnetic configurations are able to achieve such a coupling [see Fig. 2(b)]. On the other hand, the optical magnon mode dominantly couples to the phonon mode when \red{$2\varphi_\mathrm{0}\approx 70.6$ and 109.4 deg} at $\varphi_\mathrm{ext}=90(l-1)$ deg. Thus, \red{such a near-orthogonal configuration} is able to achieve a phonon-magnon coupling to the optical magnon mode. Moreover, it is obvious that a smaller mode volume, $L_\mathrm{A}\lambda$, and larger mode overlap, $L_\mathrm{M}d$, enhance the phonon-magnon coupling, implying that this coupling constant has a similar standard form as in cavity QED and optomechanics \cite{walther2006cavity,aspelmeyer2014cavity}.

\subsection{Numerical calculations}
In an actual device, the dipole-dipole interaction and in-plane uniaxial magnetic anisotropy affect the magnon properties. The dipole-dipole interaction for macrospins is derived by spatially integrating the microscopic dipole-dipole interaction over the ferromagnetic layer \cite{benson1969spin,nortemann1993microscopic}. The strength of the dipole-dipole interaction is given by $H_\mathrm{D}=M_\mathrm{S} (1-\exp[-t/\lambda])/4$, where $t$ is the thickness of each ferromagnetic layer and $\lambda$ is the wavelength of the standing spin wave. The exact expressions for the stray fields due to the self and interlayer dipole-dipole interactions are shown in Appendix A. Such a macroscopic net dipole-dipole interaction modifies the magnon dispersion and interaction with respect to the relative angle between the magnetization direction and the standing wave propagation axis. On the other hand, the in-plane uniaxial magnetic anisotropy is determined by various contributions, such as the ferromagnetic crystal properties, device geometry, and impurities in the material. Here, we assume an in-plane uniaxial magnetic anisotropy field of 0.1 mT, which is small but large enough to have a non-negligible influence in the numerical simulations. 

In the numerical investigation described below, we use several parameters for the material properties and device geometry of the three-layer CoFeB/Ru/CoFeB SAF structure on a piezoelectric $\mathrm{LiNbO}_3$ substrate that was reported in a previous magnomechanics experiment (see Table I) \cite{matsumoto2022large}. The interlayer exchange field was fixed $\mu_0H_\mathrm{E}=50$ mT. The phonon frequency shifts and the linewidth broadening given by Eqs. (\ref{freq}) and (\ref{linewidth}) were numerically calculated with respect to the strength $\mu_0H_\mathrm{ext}$ and angle $\varphi_\mathrm{ext}$ of the external magnetic field (the complete form of $A_\mathrm{M}$ is shown in Appendix E). Note that the magnetization angle under the equilibrium condition was obtained by minimizing the total energy density, $E(\varphi_1,\varphi_2)$. To avoid local minima in the minimization algorithm, the initial conditions were selected as $\varphi_{1,0}=\varphi_\mathrm{ext}+\varphi_0$ and $\varphi_{2,0}=\varphi_\mathrm{ext}-\varphi_0$, which are the exact solutions when the dipole-dipole interaction and in-plane uniaxial magnetic anisotropy are neglected. To quantitatively discuss the effect of the dipole-dipole interaction and uniaxial magnetic anisotropy, we investigate three parameter regimes: a weak dipole regime $(\mu_0H_\mathrm{D}, \mu_0H_\mathrm{uni})=(1,0)$ mT, a strong dipole regime $(\mu_0H_\mathrm{D}, \mu_0H_\mathrm{uni})=(100,0)$ mT, and an anisotropic regime $(\mu_0H_\mathrm{D}, \mu_0H_\mathrm{uni})=(50,0.1)$ mT. Note that the strength of dipolar fields $\mu_0H_\mathrm{D}=1$ (100) mT at 1 GHz phonon frequency with $\mu_0 M_\mathrm{S}=1.5$ T corresponds to the thickness of $t=9.3$ ($1.0\times10^3$) nm for which our assumption, $2t<\lambda$, holds.

\begin{table}[hbtp]
  \caption{Parameters used in numerical calculations}
  \label{table:parameters}
  \centering
  \begin{tabular}{cccc}
    \hline
    parameter &notation  & value  &  unit  \\
    \hline \hline
    Magnetoelastic coupling strength  & $b_1$  & -2 & T \\
    Gilbert damping & $\alpha$ &0.01 & no unit\\
    Gyromagnetic ratio & $\gamma$ & $1.76\times10^{11}$ &1/(s T)\\
    Magnetic saturation field & $\mu_0M_\mathrm{S}$ & 1.5 & T\\
    Total layer thickness & $d$ & $2\times 10^{-8}$ & m\\
    Density of phonon cavity  &$\rho$& 4650   & $\mathrm{kg/m^3}$ \\
    Sound velocity of phonon cavity &$v_0$& 3500 & m/s\\
    Cavity length ratio & $L_\mathrm{M}/L_\mathrm{A}$ &1/7& no unit\\
    \hline
  \end{tabular}
\end{table}

The weak dipole regime, $(\mu_0H_\mathrm{D}, \mu_0H_\mathrm{uni})=(1,0)$ mT, was examined first to confirm that the analytical expressions in Eqs. (\ref{g_ac}) and (\ref{g_opt}) are consistent. In this regime, the magnon frequency is almost independent of $\varphi_\mathrm{ext}$ and is determined by $\mu_0H_\mathrm{ext}$ [Fig. 3(a)]. The phonon frequency shift and the linewidth broadening were calculated for three different phonon frequencies [Fig. 3(b-g)]. As predicted in Eqs. (\ref{freq}) and (\ref{linewidth}), a large frequency shift and linewidth broadening occur when the magnon resonance frequency matches the phonon frequency. Moreover, the frequency shift and the linewidth broadening appear to be sinusoidal functions of $\varphi_\mathrm{ext}$. The resonance frequency of the acoustic (optical) magnon mode shows a dependence of $\sin 2\varphi_\mathrm{ext}$ ($\cos2\varphi_\mathrm{ext}$). This result is consistent with the analytical formulas in Eqs. (\ref{g_ac}) and (\ref{g_opt}).
\begin{figure*}[htbp]
  \centering
  \includegraphics[width=15cm]{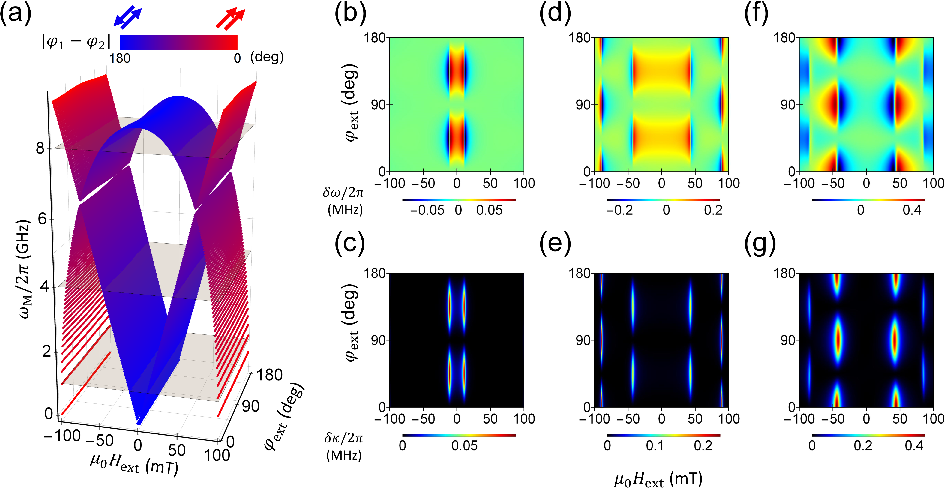}
  \caption{(a) Magnon resonance frequency $\omega_\mathrm{M}/2\pi$ as functions of the strength $\mu_0H_\mathrm{ext}$ and angle $\varphi_\mathrm{ext}$ of the external magnetic field when $(\mu_0H_\mathrm{E},\mu_0H_\mathrm{D},\mu_0H_\mathrm{uni})=(50,1,0)$ mT. The gray planes correspond to frequencies of 1 GHz, 4 GHz, and 8 GHz. The blue (red) points show the relative angle $|\varphi_1-\varphi_2|$ for the antiferromagnetic (ferromagnetic) configuration of the magnetization vectors. (b) Phonon frequency shift (b), (d), (f) and linewidth broadening (c), (e), (g) with respect to $\mu_0H_\mathrm{ext}$ and $\varphi_\mathrm{ext}$. The phonon frequency is set to be 1 GHz [(b) and (c)], 4 GHz [(d) and (e)], and 8 GHz [(f) and (g)]. }
\end{figure*}

Next, the strong dipole regime, $(\mu_0H_\mathrm{D}, \mu_0H_\mathrm{uni})=(100,0)$ mT, is investigated to reveal how the dipolar field modifies the phonon spectra. As is clearly shown in Fig. 4(a), the magnon frequency strongly depends on $\varphi_\mathrm{ext}$ because of the strong dipolar field, which is dependent on the relative angle between the magnetization direction and standing wave axis. A large energy gap appears around $\varphi_\mathrm{ext}=45$ deg and $135$ deg, indicating anti-crossing between the acoustic and optical magnon modes [see Fig. 4(b)] \cite{shiota2020tunable,sud2020tunable}. In this regime, the acoustic and optical magnon modes are hybridized and are no longer able to be represented as pure in-phase and anti-phase precessions. This can be observed as the phonon frequency shift and the linewidth broadening where the sinusoidal dependency is distorted.
\begin{figure*}[htbp]
  \centering
  \includegraphics[width=17cm]{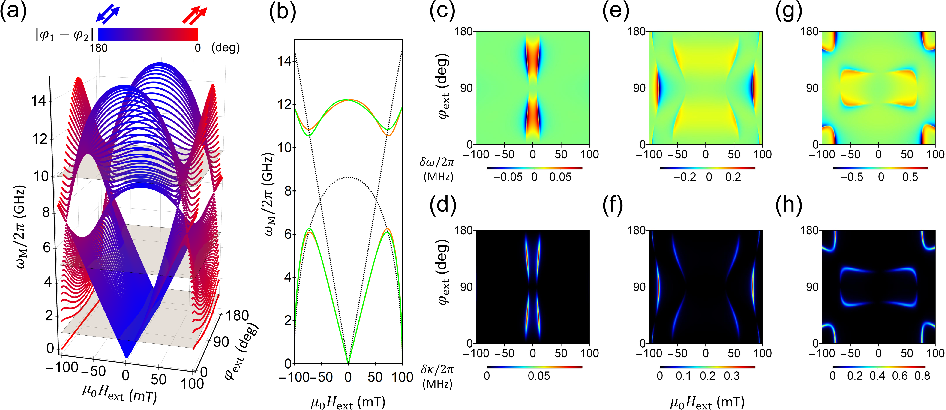}
  \caption{(a) Magnon resonance frequency $\omega_\mathrm{M}/2\pi$ as functions of the strength $\mu_0H_\mathrm{ext}$ and angle $\varphi_\mathrm{ext}$ of the external magnetic field when $(\mu_0H_\mathrm{E},\mu_0H_\mathrm{D},\mu_0H_\mathrm{uni})=(50,100,0)$ mT. The gray planes correspond to frequencies of 1 GHz, 5 GHz, and 10 GHz. The blue (red) points show the relative angle $|\varphi_1-\varphi_2|$ for the antiferromagnetic (ferromagnetic) configuration of the magnetization vectors. Because of the dipole-dipole interaction, a large energy gap appears at $\varphi_\mathrm{ext}=$ 45 deg (orange line) and 135 deg (green line) whereas there is no gap at $\varphi_\mathrm{ext}=$ 90 deg (black dashed line) (b). Phonon frequency shift (c), (e), (g) and linewidth broadening (d), (f), (h) with respect to $\mu_0H_\mathrm{ext}$ and $\varphi_\mathrm{ext}$. The phonon frequency is set to be 1 GHz [(c) and (d)], 5 GHz [(e) and (f)], and 10 GHz [(g) and (h)]. }
\end{figure*}

\begin{figure*}[htbp]
  \centering
  \includegraphics[width=12cm]{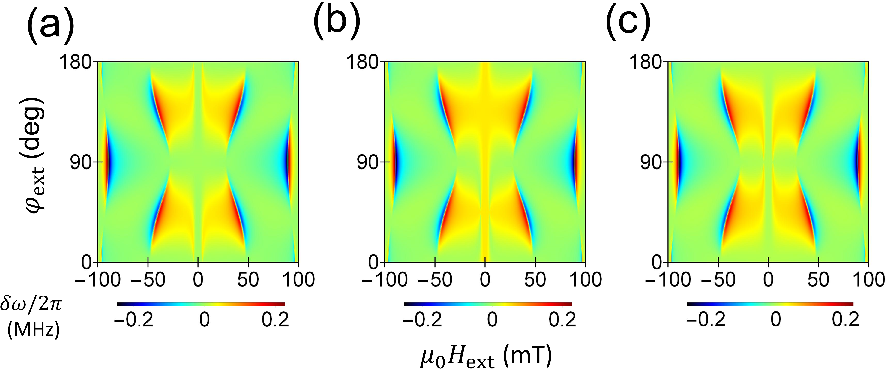}
  \caption{Phonon frequency shift at $\omega_\mathrm{P}/2\pi=4$ GHz with (a) $\varphi_\mathrm{in}=0$ deg, (b) $\varphi_\mathrm{in}=45$ deg, (c) $\varphi_\mathrm{in}=90$ deg.}
\end{figure*}
When a uniaxial magnetic anisotropy is present in the film plane, the magnetization directions align along the easy axis with the anti-parallel configuration at zero magnetic field. This feature is investigated with $(\mu_0H_\mathrm{D}, \mu_0H_\mathrm{uni})=(50,0.1)$ mT. Figure 5 shows the phonon frequency shift at $\omega_\mathrm{P}/2\pi=4$ GHz for different orientations of magnetic anisotropy, $\varphi_\mathrm{in}=0,45$, and $90$ deg. The results indicate that a small magnetic anisotropy can modify the profile of the phonon frequency shift around zero magnetic field. In particular, the result for $\varphi_\mathrm{in}=45$ deg shows relatively large frequency shifts around zero magnetic field because the magnetization angles, $\varphi_1=45$ deg and $\varphi_2=45+180$ deg, are stabilized by the in-plane uniaxial magnetic anisotropy, where the phonons efficiently couple to the acoustic magnon mode. As is the case in a SAW cavity with a single ferromagnetic layer \cite{hatanaka2022chip,hatanaka2023phononic}, the in-plane uniaxial magnetic anisotropy plays an important role in the phonon-magnon energy transduction when the external magnetic field is around zero.

\section{Discussion}
\red{A straightforward way of implementing the cavity magnomechanical system discussed above is to replace the single-layer ferromagnet by the SAF structure in an on-chip cavity magnomechanical setup \cite{hatanaka2022chip}. There are a variety of candidate SAF structures, such as CoFeB/Ru/CoFeB and Co/Ru/Co, as well as even a layered antiferromagnet like $\mathrm{CrCl_3}$, all of which have been recently utilized for investigating magnomechanics with propagating phonons \cite{matsumoto2022large, kuss2023nonreciprocal,lyons2023acoustically}. For instance, a phonon resonance frequency of 3-5 GHz is suitable for observing both acoustic and optical magnon modes in the CoFeB/Ru/CoFeB SAF structure, as shown in Fig. 4. This setup can be reasonably achieved by introducing appropriate IDTs and BRs and depositing the CoFeB/Ru/CoFeB SAF structure on a piezoelectric substrate such as lithium niobate \cite{shao2019phononic}.}

A distinct property of cavity magnomechanics with the SAF is the ability of controlling the individual coupled magnon modes by varying the strength and angle of the external magnetic field, owing to the tailorable magnetization configuration. In particular, the ability to control coupled magnon modes in the antiferromagnetic configuration is important to use antiferromagnetic dynamics for cavity magnomechanics. Our theoretical model shows that the acoustic magnon mode efficiently couples to the cavity phonon mode in the antiferromagnetic configuration. In this way, the phonon-magnon coupling is completely different from the microwave-magnon coupling \cite{shiota2020tunable}. The microwave-magnon coupling arises from a Zeeman-type energy $E_\mathrm{MW}=-\mu_0\vect{H}_\mathrm{MW}\cdot (\vect{m}_1+\vect{m}_2)$, where the optical (acoustic) magnon dominantly couples to the photon mode in the antiferromagnetic (ferromagnetic) configuration. Because the resonance frequencies of the acoustic magnon modes are generally lower than those of the optical magnon modes, SAW resonators with a relatively low frequency is suitable to the cavity magnomechanics framework. 

Moreover, the analytical formulation can help us make progress towards realization of cavity phonon-magnon polarons with coupled magnon modes where a strong coupling condition, {\it i.e.,} $g_i>\kappa_\mathrm{P},\kappa_\mathrm{M}$ ($i=$opt or ac) is required. For instance, substituting the parameters in Table I into Eqs. (\ref{g_ac2}) and (\ref{g_opt2}) gives the coupling constants of $g_\mathrm{ac}/2\pi\sim$ \red{9.0} MHz and $g_\mathrm{opt}/2\pi\sim$ \red{6.6} MHz with a phonon resonance frequency of \red{4} GHz. This can readily give $g_i/\kappa_\mathrm{P}>\red{5}$ with a reasonably large phonon cavity Q factor around 3000 \cite{hatanaka2023phononic}. On the other hand, from Eqs. (\ref{eps1}) to (\ref{eps4}), the magnon damping can be calculated as $\kappa_M\approx  M_\mathrm{S}\gamma\alpha\approx 0.4$ GHz with $M_\mathrm{S}\gg H_\mathrm{E}$. Thus, a low-loss ferromagnetic material for a SAF structure is required to realize cavity phonon-magnon polarons with coupled magnon modes. The strong coupling regime with multiple magnon modes allows us to import the cutting-edge protocols of cavity QED and cavity optomechanics, {\it e.g.}, non-reciprocal propagation with an artificial gauge field \cite{dalibard2011colloquium,bermudez2011synthetic,shen2016experimental,mathew2020synthetic}.

\section{Conclusion}
We have theoretically investigated cavity magnomechanical coupling between the cavity phonon mode and coupled magnon modes in a synthetic antiferromagnet. As well as the formalism incorporating an arbitrary dynamical strain field, we have provided a quantitative understanding by analyzing a simple setup with a SAW resonator. This theoretical formulation will help us not only to follow up on experimental findings but also to establish novel control protocols in analogy to cavity QED and cavity optomechanics with the goal of developing hybrid quantum technology \cite{lachance2019hybrid}. Furthermore, it is possible to extend this cavity magnomechanics framework to include multiple ferromagnetic layers, such as van der Waals layered crystals \cite{narath1965spin,mcguire2017magnetic,lyons2023acoustically}, paving the way for antiferromagnetic hybrid quantum systems \cite{parvini2020antiferromagnetic}.  

\section*{Acknowledgments}
The authors thank H. Yamaguchi and H. Okamoto for fruitful discussions. This work was partly supported by JSPS KAKENHI (Grant Number 20J20952, \red{23H05463}) from JSPS and JSR Fellowship from the University of Tokyo.

\clearpage
\onecolumngrid
\appendix
\section{Expression for the effective magnetic fields}
The effective magnetic field for each magnetic contribution is formulated as follows. First, we express the effective magnetic field in the $i$th layer associated with the external magnetic field, $\vect{H}_{\mathrm{ext},i}$, the demagnetization, $\vect{H}_{\mathrm{dem},i}$, the in-plane uniaxial magnetic anisotropy, $\vect{H}_{\mathrm{uni},i}$, and the interlayer exchange coupling $\vect{H}_{\mathrm{int},i}$. Each energy density is, respectively, given by
\begin{align}
E_\mathrm{ext}=&-\mu_0M_\mathrm{S} \vect{H}_\mathrm{0}\cdot(\vect{m}'_1+\vect{m}'_2),\\
E_\mathrm{dem}=& \mu_0\frac{M_\mathrm{S}^2}{2}((m'_{1z})^2+(m'_{2z})^2),\\
E_\mathrm{uni}=&-\mu_0H_\mathrm{uni}M_\mathrm{S}((\vect{m}'_1\cdot \vect{s}_\mathrm{in})^2+(\vect{m}'_2\cdot \vect{s}_\mathrm{in})^2),\\
E_\mathrm{int}=&\mu_0H_\mathrm{E}M_\mathrm{S}(\vect{m}'_1\cdot\vect{m}'_2),
\end{align}
where $\vect{H}_\mathrm{0}=H_\mathrm{ext}(\cos\varphi_\mathrm{ext},\sin\varphi_\mathrm{ext},0)$ is the in-plane external magnetic field with the field angle $\varphi_\mathrm{ext}$ with respect to the $X$ axis, $H_\mathrm{uni}$ is the strength of the in-plane uniaxial magnetic anisotropy, $\vect{s}_\mathrm{in}=(\cos\varphi_\mathrm{in},\sin\varphi_\mathrm{in},0)$ is the easy axis of the in-plane uniaxial magnetic anisotropy with the anisotropy angle $\varphi_\mathrm{in}$, $H_\mathrm{E}$ is the interlayer exchange coupling field, and $\vect{m}'_i$ is the magnetization vector in the laboratory frame. By bringing them into the spin frame, we obtain
\begin{align}
E_\mathrm{ext}=&-\mu_0M_\mathrm{S}H_\mathrm{ext}\sum_{i=1,2}\left[\cos(\varphi_\mathrm{ext}-\varphi_i)+m_{iy}\sin(\varphi_\mathrm{ext}-\varphi_i)\right],\\
E_\mathrm{dem}=&\mu_0\frac{M^2_\mathrm{S}}{2}(m_{1x}^2+m_{2x}^2),\\
E_\mathrm{uni}=&-\mu_0H_\mathrm{uni}\sum_{i=1,2}\left[\cos(\varphi_i-\varphi_\mathrm{in})-m_{iy}\sin(\varphi_i-\varphi_\mathrm{in})\right]^2,\\
E_\mathrm{int}=&\mu_0H_\mathrm{E}M_\mathrm{S}\left[\cos(\varphi_1-\varphi_2)-m_{1y}\sin(\varphi_1-\varphi_2)+m_{2y}\sin(\varphi_1-\varphi_2)\right].
\end{align}
Thus, the effective magnetic field becomes
\begin{align}
\vect{H}_{\mathrm{ext},i}=&(0,H_\mathrm{ext}\sin(\varphi_\mathrm{ext}-\varphi_i),0),\\
\vect{H}_{\mathrm{dem},i}=&(-M_\mathrm{S} m_{ix}/2 ,0,0),\\
\vect{H}_{\mathrm{uni},i}=&(0,-2H_\mathrm{uni}(\cos(\varphi_i-\varphi_\mathrm{in})-m_{iy}\sin(\varphi_i-\varphi_\mathrm{in})),0),\\
\vect{H}_{\mathrm{int},i}=&-H_\mathrm{E}(m_{\bar{i}x},m_{\bar{i}y}\cos(\varphi_1-\varphi_2)+(-1)^i \sin(\varphi_1-\varphi_2),0),
\end{align}
where $\bar{i}$ is the opposite layer index for $i$. 

The effective magnetic field from the dipole-dipole interaction can be formulated by integrating the microscopic dipolar field over the ferromagnetic layer in a similar way to what is shown in Refs \cite{benson1969spin,nortemann1993microscopic}. Here, we emphasize that the dipole-dipole interaction in our model is calculated by the standing wave expansion because our assumption is that the standing phonon wave couples to the standing spin wave. In the long wavelength limit, the dipole sums $D_{ij}$ defined in \cite{benson1969spin} have the form,
\begin{align}
D_{YY}=&H_\mathrm{D}\cos^2\varphi_d,\\
D_{XX}=&-H_\mathrm{D},\\
D_{ZZ}=&H_\mathrm{D}\sin^2\varphi_d ,\\
D_{YZ}=&D_{ZY}=H_\mathrm{D}\sin\varphi_d \cos\varphi_d,\\
D_{XZ}=&D_{ZX}=D_{XY}=D_{YX}=0,
\end{align}
where $\varphi_\mathrm{d}$ shows the angle between the spin wave propagation direction and the $X$ axis. The dipole magnetic field $H_\mathrm{D}=M_\mathrm{S}(1- \exp[-kt])/4$ is determined by the thickness of the ferromagnetic layer, $t$, and the spin wavevector $k$. The self-dipolar field, which is exerted on the $i$th layer by the dipole-dipole interaction, is given by
\begin{align}
\vect{H}_\mathrm{self,i}=-H_\mathrm{D} \left(\begin{matrix}
-1&0&0\\
0&\cos^2\varphi_{di}&\sin\varphi_{di}\cos\varphi_{di}\\
0&\sin\varphi_{di}\cos\varphi_{di}&\sin^2\varphi_{di}
\end{matrix}\right) \left(\begin{matrix}
m_{ix}\\m_{iy}\\m_{iz}
\end{matrix}\right)=H_\mathrm{D}\left(\begin{matrix}
m_{ix}\\
-m_{iy}\cos^2\varphi_{di}\\
-m_{iy}\sin\varphi_{di}\cos\varphi_{di}
\end{matrix}\right)=H_\mathrm{D}\left(\begin{matrix}
m_{ix}\\
-m_{iy}\sin^2\varphi_{i}\\
-m_{iy}\sin\varphi_{i}\cos\varphi_{i}
\end{matrix}\right),
\end{align}
Here, we note that $\varphi_{di}=\pi/2-\varphi_i$ by taking into account the acoustic wave along the $X$ direction. In the same manner, the interlayer dipolar field, which is exerted on the $i$th layer due to the dipole-dipole interaction in the $j$th layer, is given by
\begin{align}
\vect{H}_\mathrm{j\to i}=&-H_\mathrm{D}\mathcal{R}(-\varphi_i)\mathcal{R}(\varphi_j)\left(\begin{matrix}
-1&0&0\\
0&\cos^2\varphi_{dj}&\sin\varphi_{dj}\cos\varphi_{dj}\\
0&\sin\varphi_{dj}\cos\varphi_{dj}&\sin^2\varphi_{dj}
\end{matrix}\right) \left(\begin{matrix}
m_{jx}\\m_{jy}\\m_{jz}
\end{matrix}\right)=H_\mathrm{D}\left(\begin{matrix}
m_{jx}\\
-m_{jy}\cos\varphi_{dj}\cos(\varphi_i-\varphi_j-\varphi_{dj})\\
m_{jy}\cos\varphi_{dj}\sin(\varphi_i-\varphi_j-\varphi_{dj})
\end{matrix}\right)\nonumber\\
&=H_\mathrm{D}\left(\begin{matrix}
m_{jx}\\
-m_{jy}\sin\varphi_i \sin\varphi_j\\
-m_{jy}\cos\varphi_i \sin\varphi_j
\end{matrix}\right),
\end{align}
where $\mathcal{R}(\varphi)$ shows the in-plane rotation matrix with the angle of $\varphi$. Thus, the effective magnetic field from the dipole-dipole interaction is given by $\vect{H}_{\mathrm{dip},i}=\vect{H}_{\mathrm{self},i}+\vect{H}_{j\to i}$.

Finally, the effective magnetic field from the magnetoelastic coupling is calculated from the energy density,
\begin{align}
E_\mathrm{PM}=\mu_0M_\mathrm{S}\sum_{i=1,2}\sum_{j,k={x,y,z}} b^{(i)}_{jk} \varepsilon_{jk} m'_{ij} m'_{ik},
\end{align}
where $b^{(i)}_{jk}$ is the magnetoelastic coupling strength in the $i$th ferromagnetic layer, and $\varepsilon_{jk}$ is the strain field. Note that the magnetoelastic coupling strength is typically $b^{(i)}_{jk}=b^{(i)}_1$ when $j=k$ for the longitudinal dynamic magnetostriction and $b^{(i)}_{jk}=b^{(i)}_2$ when $j\neq k$ for the shear dynamic magnetostriction. Bringing it into the spin frame, it can be linearized by performing a linear precession approximation as
\begin{align}
E_\mathrm{PM}=\mu_0M_\mathrm{S}\sum_{i=1,2}\left[2b^{(i)}_2\left(\varepsilon_{yz}m_{ix}\sin\varphi_i+\varepsilon_{xz}m_{ix}\cos\varphi_i-\varepsilon_{xy}\cos2\varphi_i\right)+b^{(i)}_1\left(\varepsilon_{yy}-\varepsilon_{xx}\right)m_{iy}\sin2\varphi_i\right].
\end{align}
Thus, the effective magnetic field from the magnetoelastic coupling becomes
\begin{align}
\vect{H}_{\mathrm{PM},i}=(2b^{(i)}_2(\varepsilon_{xz}\cos\varphi_i+\varepsilon_{yz}\sin\varphi_i),-2b^{(i)}_2\varepsilon_{xy}\cos2\varphi_i+b^{(i)}_1(\varepsilon_{xx}-\varepsilon_{yy})\sin2\varphi_i,0).
\end{align}

\section{Expression for $\vect{q}$}
In the linearized LLG equation, the four-dimensional vector $\vect{b}_\mathrm{M}$ is expressed as
\begin{align}
\vect{b}_\mathrm{M}=& D_\mathrm{M}\left(
\begin{array}{c}
2b^{(1)}_2\varepsilon_{xy}\cos2\varphi_1-b^{(1)}_1(\varepsilon_{xx}-\varepsilon_{yy})\sin2\varphi_1\\
2b^{(1)}_2\left(\varepsilon_{xz}\cos\varphi_1+\varepsilon_{yz}\sin\varphi_1\right)\\
2b^{(2)}_2\varepsilon_{xy}\cos2\varphi_2-b^{(2)}_1(\varepsilon_{xx}-\varepsilon_{yy})\sin2\varphi_2\\
2b^{(2)}_2\left(\varepsilon_{xz}\cos\varphi_2+\varepsilon_{yz}\sin\varphi_2\right)
\end{array}
\right)\nonumber\\
=&\varepsilon_{xx}\vect{q}_{xx}+\varepsilon_{yy}\vect{q}_{yy}+\varepsilon_{xy}\vect{q}_{xy}+\varepsilon_{yz}\vect{q}_{yz}+\varepsilon_{zx}\vect{q}_{zx},
\end{align}
where
\begin{align}
D_\mathrm{M}=-\gamma\left(\begin{array}{cccc}
1&\alpha&0&0\\
-\alpha&1&0&0\\
0&0&1&\alpha\\
0&0&-\alpha&1
\end{array}\right)^{-1}=-\frac{\gamma}{1+\alpha^2}\left(\begin{array}{cccc}
1&-\alpha&0&0\\
\alpha&1&0&0\\
0&0&1&-\alpha\\
0&0&\alpha&1
\end{array}\right).
\end{align}
Thus, we obtain
\begin{align}
\vect{q}_{xx}=&-\vect{q}_{yy}=-\frac{\gamma}{1+\alpha^2}\left(b^{(1)}_1\sin2\varphi_1,b^{(1)}_1\alpha\sin2\varphi_1,b^{(2)}_1\sin2\varphi_2,-b^{(2)}_{1}\alpha\sin2\varphi_2\right)^\mathrm{T},\label{SAWqxx}\\
\vect{q}_{xy}=&\frac{2\gamma}{1+\alpha^2}\left(b^{(1)}_2\alpha\cos2\varphi_1,b^{(1)}_2\alpha\cos2\varphi_1,b^{(2)}_2\cos2\varphi_2,b^{(2)}_2\alpha\cos2\varphi_2\right)^\mathrm{T},\\
\vect{q}_{yz}=&\frac{2\gamma}{1+\alpha^2}\left(b^{(1)}_2\alpha \sin\varphi_1,-b^{(1)}_2 \sin\varphi_1, b^{(2)}_2\alpha \sin\varphi_2,-b^{(2)}_2 \sin\varphi_2,\right)^\mathrm{T},\\
\vect{q}_{zx}=&\frac{2\gamma}{1+\alpha^2}\left(b^{(1)}_2\alpha \cos\varphi_1,-b^{(1)}_2 \cos\varphi_1, b^{(2)}_2 \alpha\cos\varphi_2,-b^{(2)}_2 \cos\varphi_2 \right)^\mathrm{T}.
\end{align}
In case of the SAW resonator setup, only $\vect{q}_{xx}$ has to be considered.

\section{Expression for $\bar{\vect{q}}$}
Regarding the expression for $\bar{\vect{q}}$, we explicitly show the inner product in the dynamic magnetostrictive force in Eq. (\ref{EQM1}),
\begin{align}
\vect{\psi}\cdot\vect{f}_\mathrm{PM}=\sum_{i=1,2}\Biggl[&\psi_X\left(b^{(i)}_2\partial_Z m_{ix}\cos\varphi_i+b^{(i)}_2\partial_Y m_{iy}\cos2\varphi_i-b^{(i)}_1 \partial_X m_{iy}\sin2\varphi_i\right)\nonumber\\
&+\psi_Y\left(b^{(i)}_2 \partial_X m_{iy}\cos2\varphi_i+b^{(i)}_2\partial_Z m_{ix}\sin\varphi_i+b^{(i)}_1\partial_Y m_{iy}\sin2\varphi_i\right)\nonumber\\
&+\psi_Z\left(b^{(i)}_2\partial_X m_{ix}\cos\varphi_i+b^{(i)}_2\partial_Y m_{ix}\sin\varphi   \right)\Biggr].
\end{align}
This can be further decomposed into $\vect{\psi}\cdot\vect{f}_\mathrm{PM}=\bar{\vect{q}}\cdot P\tilde{\vect{m}}$ with the four-dimensional vector operator $\bar{\vect{q}}$,
\begin{align}
\bar{\vect{q}}=&\left(\begin{array}{c}
\psi_Xb^{(1)}_2\cos\varphi_1 \partial_Z+\psi_Y b^{(1)}_2 \sin\varphi_1 \partial_Z+\psi_Zb^{(1)}_2(\cos\varphi_1 \partial_X+\sin\varphi_1 \partial_Y) \\
\psi_X(b^{(1)}_2\cos2\varphi_1 \partial_Y-b^{(1)}_1 \sin2\varphi_1 \partial_X)+\psi_Y(b^{(1)}_2 \cos2\varphi_1 \partial_X+b^{(1)}_1\sin2\varphi_1\partial_Y) \\
\psi_Xb^{(2)}_2\cos\varphi_2 \partial_Z+\psi_Y b^{(2)}_2 \sin\varphi_2 \partial_Z+\psi_Zb^{(2)}_2(\cos\varphi_2 \partial_X+\sin\varphi_2 \partial_Y) \\
\psi_X(b^{(2)}_2\cos2\varphi_2 \partial_Y-b^{(2)}_1 \sin2\varphi_2 \partial_X)+\psi_Y(b^{(2)}_2 \cos2\varphi_2 \partial_X+b^{(2)}_1\sin2\varphi_2\partial_Y) 
\end{array}\right).
\end{align}
In case of the SAW resonator setup, we can simplify the above equation by setting $\psi_Y=\psi_Z=0$ and $\partial_Y=\partial_Z=0$:
\begin{align}
\bar{\vect{q}}=&\left(\begin{array}{c}
0\\
-\psi_Xb^{(1)}_1 \sin2\varphi_1 \partial_X\\
0\\
-\psi_X b^{(2)}_1 \sin2\varphi_2 \partial_X
\end{array}\right).\label{SAWqbar}
\end{align}

\section{Analytical derivation of magnomechanical coupling constants}
We start from the linearized LLG equation with the matrix $A_\mathrm{M}$ given in Eq. (\ref{SAW_AM}), which can be written as
\begin{align}
A_\mathrm{M}=-\frac{\gamma}{1+\alpha^2}\left(\begin{array}{cccc}
-(H_\mathrm{E}+ M_\mathrm{S})\alpha&-H_\mathrm{E}&-H_\mathrm{E}\alpha&-a_c\\
H_\mathrm{E}+M_\mathrm{S}&-H_\mathrm{E}\alpha&H_\mathrm{E}&-a_c\alpha\\
-H_\mathrm{E}\alpha&-a_c&-(H_\mathrm{E}+ M_\mathrm{S})\alpha&-H_\mathrm{E}\\
H_\mathrm{E}&-a_c\alpha&H_\mathrm{E}+ M_\mathrm{S}&-H_\mathrm{E}\alpha
\end{array}\right).
\end{align}
The eigenvalues are calculated as
\begin{align}
\epsilon_{1}=&\frac{\gamma}{2(1+\alpha^2)}\left[-2H_\mathrm{E}\alpha\sin^2\varphi_0- M_\mathrm{S}\alpha+\sqrt{-8 M_\mathrm{S} H_\mathrm{E}\sin^2\varphi_0+\alpha^2\left(2H_\mathrm{E}\sin^2\varphi_0- M_\mathrm{S} \right)^2}\right],\\
\epsilon_{2}=&\frac{\gamma}{2(1+\alpha^2)}\left[-2H_\mathrm{E}\alpha\sin^2\varphi_0-M_\mathrm{S}\alpha-\sqrt{-8 M_\mathrm{S} H_\mathrm{E}\sin^2\varphi_0+\alpha^2\left(2H_\mathrm{E}\sin^2\varphi_0- M_\mathrm{S} \right)^2}\right],
\\
\epsilon_{3}=&\frac{\gamma}{2(1+\alpha^2)}\left[-H_\mathrm{E}\alpha(\cos2\varphi_0+3)- M_\mathrm{S}\alpha+\sqrt{-8H_\mathrm{E}(2H_\mathrm{E}+ M_\mathrm{S})\cos^2\varphi_0+\alpha^2\left(2H_\mathrm{E}\sin^2\varphi_0+M_\mathrm{S} \right)^2}\right],\\
\epsilon_{4}=&\frac{\gamma}{2(1+\alpha^2)}\left[-H_\mathrm{E}\alpha(\cos2\varphi_0+3)-M_\mathrm{S}\alpha-\sqrt{-8H_\mathrm{E}(2H_\mathrm{E}+M_\mathrm{S})\cos^2\varphi_0+\alpha^2\left(2H_\mathrm{E}\sin^2\varphi_0+M_\mathrm{S} \right)^2}\right].
\end{align}
Because typical ferromagnetic materials show $\alpha<1$, they can be approximated to first order of $\alpha$ as follows:
\begin{align}
\epsilon_{1}=&-i\gamma\sqrt{2 H_\mathrm{E}M_\mathrm{S}}\red{|}\sin\varphi_0\red{|}-\frac{\gamma}{2}\alpha(2H_\mathrm{E}\sin^2\varphi_0+M_\mathrm{S}) +\mathcal{O}(\alpha^2),\label{eps1}\\
\epsilon_{2}=&i\gamma\sqrt{2 H_\mathrm{E}M_\mathrm{S}}\red{|}\sin\varphi_0\red{|}-\frac{\gamma}{2}\alpha(2H_\mathrm{E}\sin^2\varphi_0+ M_\mathrm{S}) +\mathcal{O}(\alpha^2),
\\
\epsilon_{3}=&i\gamma\sqrt{2H_\mathrm{E}(2H_\mathrm{E}+ M_\mathrm{S})}\red{|}\cos\varphi_0\red{|}-\frac{\gamma}{2}\alpha(2H_\mathrm{E}+M_\mathrm{S}+H_\mathrm{E}\cos^2\varphi_0) +\mathcal{O}(\alpha^2),\label{eigf}\\
\epsilon_{4}=&-i\gamma\sqrt{2H_\mathrm{E}(2H_\mathrm{E}+M_\mathrm{S})}\red{|}\cos\varphi_0\red{|}-\frac{\gamma}{2}\alpha(2H_\mathrm{E}+ M_\mathrm{S}+H_\mathrm{E}\cos^2\varphi_0) +\mathcal{O}(\alpha^2).\label{eps4}
\end{align}
Apparently, $\epsilon_1$ and $\epsilon_2$ ($\epsilon_3$ and $\epsilon_4$) show the energy of the same type of the coupled magnon modes. The corresponding eigenvectors can be expanded on the order of $\alpha$ as
\begin{align}
\vect{v}_1=&(v_1,-1,-v_1,1)^T, \label{eigv1}\\
\vect{v}_2=&(v_2,-1,-v_2,1)^T,\\
\vect{v}_3=&(v_3,1,v_3,1)^T,\\
\vect{v}_4=&(v_4,1,v_4,1)^T,\label{eigv4}\\
v_1=&i\sqrt{\frac{2H_\mathrm{E}}{ M_\mathrm{S}}}\red{|}\sin\varphi_0\red{|}+\frac{M_\mathrm{S}-2H_\mathrm{E}\sin^2\varphi_0}{2M_\mathrm{S}}\alpha +\mathcal{O}(\alpha^2),\\
v_2=&-i\sqrt{\frac{2H_\mathrm{E}}{ M_\mathrm{S}}}\red{|}\sin\varphi_0\red{|}+\frac{M_\mathrm{S}-2H_\mathrm{E}\sin^2\varphi_0}{2 M_\mathrm{S}}\alpha +\mathcal{O}(\alpha^2),\\
v_3=&i\sqrt{\frac{2H_\mathrm{E}}{2H_\mathrm{E}+M_\mathrm{S}}}\red{|}\cos\varphi_0\red{|}-\frac{2H_\mathrm{E}\sin^2\varphi_0+ M_\mathrm{S}}{4H_\mathrm{E}+2M_\mathrm{S}}\alpha +\mathcal{O}(\alpha^2),\\
v_4=&-i\sqrt{\frac{2H_\mathrm{E}}{2H_\mathrm{E}+ M_\mathrm{S}}}\red{|}\cos\varphi_0\red{|}-\frac{2H_\mathrm{E}\sin^2\varphi_0+M_\mathrm{S}}{4H_\mathrm{E}+ 2M_\mathrm{S}}\alpha +\mathcal{O}(\alpha^2).
\end{align} 
From the phase of the eigenvectors $\vect{v}_\mu$ $(\mu=1,2,3,4)$ in Eqs. (\ref{eigv1}) to (\ref{eigv4}), we can classify the magnon modes into two types: optical magnon modes $(\mu=1,2)$ where the phases of the magnon precessions are inverted between the ferromagnetic layers and acoustic magnon modes $(\mu=3,4)$ that have the same phase. From Eq. (\ref{int1}), we can calculate one part of the magnetoelastic coupling strength. First, we evaluate the spatial integral part with $\Phi_\mu=F(Z)\sin kX$ ($\mu=1,2,3,4$) and $\psi_X=F(Z)\cos kX$ as
\begin{align}
\frac{\int\mathrm{d}^3\vect{r} \Phi_\mu \Psi_{XX}}{\int\mathrm{d}^3\vect{r} \Phi^2_\mu}=k\frac{\int_{-L_\mathrm{M}/2}^{L_\mathrm{M}/2}\mathrm{d}X\int_0^d \mathrm{d}Z F(Z)\sin^2 kX}{\int_{-L_\mathrm{A}/2}^{L_\mathrm{A}/2}\mathrm{d}X\int_0^d \mathrm{d}Z F(Z)\sin^2 kX}\approx k,
\end{align}
where $F(Z)$ shows the phonon spatial distribution along the $Z$ axis where the internal phonon distribution depth is about $\lambda$ and the distribution depth of the ferromagnetic layer is given by its thickness $d$ with the $d\ll \lambda$ condition. Thus, the spatial integrals in the numerator and denominator are equivalent. Accordingly, by using Eq (\ref{SAWqxx}) ($b^{(1)}_1=b^{(2)}_1=b_1$), we can represent $P^{-1}\vect{q}_{xx}$ as
\begin{align}
P^{-1}\vect{q}_{xx}=\red{\frac{\gamma b_1}{1+\alpha^2}\left(\begin{matrix}
-\frac{1+v_2\alpha}{v_1-v_2}\cos2\varphi_\mathrm{ext}\sin2\varphi_0\\
\frac{1+v_1\alpha}{v_1-v_2}\cos2\varphi_\mathrm{ext}\sin2\varphi_0\\
\frac{1-v_4\alpha}{v_3-v_4}\sin2\varphi_\mathrm{ext}\cos2\varphi_0\\
-\frac{1-v_3\alpha}{v_3-v_4}\sin2\varphi_\mathrm{ext}\cos2\varphi_0
\end{matrix}\right).}
\end{align}
Thus, we obtain
\begin{align}
g_{\mathrm{MP},1}=&\red{-k\gamma b_1\frac{1+v_2\alpha}{(v_1-v_2)(1+\alpha^2)}\cos2\varphi_\mathrm{ext}\sin2\varphi_0,}\label{gMP1}\\
g_{\mathrm{MP},2}=&\red{k\gamma b_1\frac{1+v_1\alpha}{(v_1-v_2)(1+\alpha^2)}\cos2\varphi_\mathrm{ext}\sin2\varphi_0,}\\
g_{\mathrm{MP},3}=&\red{-k\gamma b_1\frac{1-v_4\alpha}{(v_3-v_4)(1+\alpha^2)}\sin2\varphi_\mathrm{ext}\cos2\varphi_0,}\\
g_{\mathrm{MP},4}=&\red{k\gamma b_1\frac{1-v_3\alpha}{(v_3-v_4)(1+\alpha^2)}\sin2\varphi_\mathrm{ext}\cos2\varphi_0.}\label{gMP4}
\end{align}
Here, we note that \red{
\begin{align}
\eta_1\equiv& -\frac{1+v_2\alpha}{(v_1-v_2)(1+\alpha^2)}=\frac{i}{2|\sin\varphi_0|}\sqrt{\frac{M_\mathrm{S}}{2H_\mathrm{E}}}+\frac{\alpha}{2}+\mathcal{O}(\alpha^2),\\
\eta_2\equiv&\frac{1+v_1\alpha}{(v_1-v_2)(1+\alpha^2)}=-\frac{i}{2|\sin\varphi_0|}\sqrt{\frac{M_\mathrm{S}}{2H_\mathrm{E}}}+\frac{\alpha}{2}+\mathcal{O}(\alpha^2),\\
\eta_3\equiv&\frac{1-v_4\alpha}{(v_3-v_4)(1+\alpha^2)}=\frac{i}{2|\cos\varphi_0|}\sqrt{\frac{2H_\mathrm{E}+M_\mathrm{S}}{2H_\mathrm{E}}}-\frac{\alpha}{2}+\mathcal{O}(\alpha^2),\\
\eta_4\equiv&-\frac{1-v_3\alpha}{(v_3-v_4)(1+\alpha^2)}=-\frac{i}{2|\cos\varphi_0|}\sqrt{\frac{2H_\mathrm{E}+M_\mathrm{S}}{2H_\mathrm{E}}}-\frac{\alpha}{2}+\mathcal{O}(\alpha^2).
\end{align}
The terms of zeroth order in $\alpha$ in $\eta_\lambda$ show the ratio between the in-plane and the out-of-components of the eigenvectors. Because the strain field in the SAW resonator is coupled to the in-plane component in this model, $g_{\mathrm{MP},i}$ is enhanced when the in-plane component is larger than the out-of-plane component.}
In a similar manner, we can calculate the other part of the magnomechanical interaction strengths in Eq.(\ref{int2}). From Eq. (\ref{SAWqbar}), we have
\begin{align}
\frac{M_\mathrm{S}\int\mathrm{d}^3\vect{r}\bar{\vect{q}}\cdot P\tilde{\vect{m}}}{\rho\int\mathrm{d}^3\vect{r}|\psi_X(\vect{r})|^2}=&-\frac{b_1kM_\mathrm{S}}{\rho}\frac{L_\mathrm{A}}{L_\mathrm{M}}\frac{d}{2\lambda}\left(\begin{matrix}0\\\sin2\varphi_1\\0\\\sin\varphi_2\end{matrix}\right)\cdot P \tilde{\vect{m}}\nonumber\\
=&-\frac{b_1kM_\mathrm{S}}{\rho}\frac{L_\mathrm{A}}{L_\mathrm{M}}\frac{d}{\lambda} \left[\cos2\varphi_\mathrm{ext}\sin2\varphi_0(-\tilde{m}_1+\tilde{m}_2)+\sin2\varphi_\mathrm{ext}\cos2\varphi_0(\tilde{m}_3+\tilde{m}_4)\right].
\end{align}
\red{Here, we should emphasize that the spatial distribution of the $P\tilde{\vect{m}}$ is characterized by $d/2\lambda$ instead of $d/\lambda$ because it is not the distribution of the coupled modes but rather that of the individual magnon modes in each ferromagnetic layer.} Thus, we obtain
\begin{align}
g_{\mathrm{PM},1}=&\frac{b_1kM_\mathrm{S}}{\rho}\frac{L_\mathrm{M}}{L_\mathrm{A}}\frac{d}{\lambda}\cos2\varphi_\mathrm{ext}\sin2\varphi_0,\label{gPM1}\\
g_{\mathrm{PM},2}=&-\frac{b_1kM_\mathrm{S}}{\rho}\frac{L_\mathrm{M}}{L_\mathrm{A}}\cos2\varphi_\mathrm{ext}\sin2\varphi_0,\label{gPM2}\\
g_{\mathrm{PM},3}=&-\frac{b_1kM_\mathrm{S}}{\rho}\frac{L_\mathrm{M}}{L_\mathrm{A}}\frac{d}{\lambda}\sin2\varphi_\mathrm{ext}\cos2\varphi_0,\label{gPM3}\\
g_{\mathrm{PM},4}=&-\frac{b_1kM_\mathrm{S}}{\rho}\frac{L_\mathrm{M}}{L_\mathrm{A}}\frac{d}{\lambda}\sin2\varphi_\mathrm{ext}\cos2\varphi_0.\label{gPM4}
\end{align}

 Importantly, the limit $\alpha\to 0$ provides us with an exact formulation of the coupling constants for the acoustic and optical magnon modes using Eqs. (\ref{gMP1}) to (\ref{gMP4}) and Eqs. (\ref{gPM1}) to (\ref{gPM4}), as follows:
\begin{align}
g_\mathrm{ac}=&\sqrt{\frac{g_\mathrm{PM,3}g_\mathrm{MP,3}}{2\omega_\mathrm{P}}}=k|b_1\sin2\varphi_\mathrm{ext}\cos2\varphi_0|\sqrt{\frac{\gamma M_\mathrm{S}\red{|\eta_3|}}{2\rho\omega_\mathrm{P}}\frac{L_\mathrm{M}}{L_\mathrm{A}}\frac{d}{\lambda}}\nonumber\\
\approx&\red{ k\left|b_1\sin2\varphi_\mathrm{ext}\frac{\cos2\varphi_0}{\sqrt{|\cos \varphi_0|}}\right|\sqrt[4]{\frac{2H_\mathrm{E}+M_\mathrm{S}}{2H_\mathrm{E}}}\sqrt{\frac{\gamma M_\mathrm{S}}{4\rho\omega_\mathrm{P}}\frac{L_\mathrm{M}}{L_\mathrm{A}}\frac{d}{\lambda}} },\\
g_\mathrm{opt}=&\sqrt{\frac{g_\mathrm{PM,1}g_\mathrm{MP,1}}{2\omega_\mathrm{P}}}=k|b_1\cos2\varphi_\mathrm{ext}\sin2\varphi_0|\sqrt{\frac{\gamma M_\mathrm{S}\red{|\eta_1|}}{2\rho\omega_\mathrm{P}}\frac{L_\mathrm{M}}{L_\mathrm{A}}\frac{d}{\lambda}}\nonumber\\
\approx&\red{k\left|b_1\cos2\varphi_\mathrm{ext}\frac{\sin2\varphi_0}{\sqrt{|\sin\varphi_0|}}\right|\sqrt[4]{\frac{M_\mathrm{S}}{2H_\mathrm{E}}}\sqrt{\frac{\gamma M_\mathrm{S}}{4\rho\omega_\mathrm{P}}\frac{L_\mathrm{M}}{L_\mathrm{A}}\frac{d}{\lambda}}.}
\end{align}
Factors of $L_\mathrm{A}/L_\mathrm{M}$ and $d/\lambda$ appear as the ratio between the mode overlap of the phonon and magnon modes and the mode volume of the phonon mode.

\red{The magnetoelastic coupling constant for the acoustic magnon, $g_\mathrm{ac}$, shows a relative angular dependence of $|\cos2\varphi_0/\sqrt{|\cos\varphi_0|}|$ that becomes zero at $2\varphi_0=90$ deg (orthogonal configuration) and unity at $2\varphi_0=0$ deg (ferromagnetic configuration). On the other hand, it diverges at $2\varphi_0=180$ deg because the out-of-plane component of the eigenvector becomes zero. Here, we should emphasize that the condition $2\varphi_0=180$ ({\it i.e.}, $H_\mathrm{ext}=0$) never shows a physically meaningful coupling in this model because of the zero eigen frequency of the magnon modes [see Eq. (\ref{eigf})]. A physically meaningful and experimentally accessible definition of $g^\mathrm{res}_\mathrm{ac}$ is given by taking the resonance condition ($\omega_\mathrm{P}$=$\omega_\mathrm{M}=\mathrm{Im}[\epsilon_3]$) from Eq. (\ref{eigf}). It can be represented as
\begin{align}
g^\mathrm{res}_\mathrm{ac}=& \red{k\left|b_1\sin2\varphi_\mathrm{ext}\right|\frac{\gamma M_\mathrm{S}}{\omega_\mathrm{P}}\sqrt{\frac{\mu_0}{4\rho}\frac{L_\mathrm{M}}{L_\mathrm{A}}\frac{d}{\lambda}} }\label{g_ac2},
\end{align}
where we assume that $M_\mathrm{S}\gg H_\mathrm{E}$.}

\red{The magnetoelastic coupling constant for the optical magnon, $g_\mathrm{opt}$, takes a relative angle dependence of $|\sin2\varphi_0/\sqrt{|\sin\varphi_0|}|$ that becomes zero at $2\varphi_0=180$ deg (antiferromagnetic configuration) and 0 deg (ferromagnetic configuration). Moreover, it takes the maximum values of $|\sin2\varphi_0/\sqrt{|\sin\varphi_0|}|\approx 1.2$ at $2\varphi_0\approx 70.6$ deg and $109.4$ deg. Thus, it can be represented as
\begin{align}
\max_{\varphi_0} g_\mathrm{opt}\approx&\red{1.2 k\left|b_1\cos2\varphi_\mathrm{ext}\right|\sqrt[4]{\frac{M_\mathrm{S}}{H_\mathrm{E}}}\sqrt{\frac{\gamma M_\mathrm{S}}{4\rho\omega_\mathrm{P}}\frac{L_\mathrm{M}}{L_\mathrm{A}}\frac{d}{\lambda}} }\label{g_opt2}.
\end{align}
In the same manner for the resonance condition, $g^\mathrm{res}_\mathrm{opt}$ is given by
\begin{align}
g^\mathrm{res}_\mathrm{opt}=&\red{k\left|b_1\cos2\varphi_\mathrm{ext}\right|\sqrt{\frac{M_\mathrm{S}}{\mu_0 H_\mathrm{E}}}\sqrt{\frac{1}{2\rho}\frac{L_\mathrm{M}}{L_\mathrm{A}}\frac{d}{\lambda}} }\label{g_opt2}.
\end{align}
}

\section{Expression for $A_\mathrm{M}$}
It is helpful to explicitly write down the exact form of the matrix $A$ appearing in the linearized LLG equation $A_\mathrm{M}=\mu_0 D_\mathrm{M}A$ with the full effective magnetic field in Eq. (\ref{effH}), as follows:
\begin{align}
A_{11}=&0,\\
A_{12}=&H_\mathrm{ext}\cos(\varphi_1-\varphi_\mathrm{ext})-H_\mathrm{E}\cos(\varphi_1-\varphi_2)+2H_\mathrm{uni} \cos(2(\varphi_1-\varphi_\mathrm{in}))+H_\mathrm{D}\sin^2\varphi_1,\\
A_{13}=&0,\\
A_{14}=&H_\mathrm{E}\cos(\varphi_1-\varphi_2)+H_\mathrm{D} \sin\varphi_1 \sin\varphi_2,\\
A_{21}=&-M_\mathrm{S} -H_\mathrm{ext}\cos(\varphi_1-\varphi_\mathrm{ext})+H_\mathrm{E}\cos(\varphi_1-\varphi_2)-2H_\mathrm{uni}\cos^2(\varphi_1-\varphi_\mathrm{in})+H_\mathrm{D},\\
A_{22}=&0,\\
A_{23}=&-H_\mathrm{E}+H_\mathrm{D},\\
A_{24}=&0,\\
A_{31}=&0,\\
A_{32}=&H_\mathrm{E}\cos(\varphi_1-\varphi_2)+H_\mathrm{D}\sin\varphi_1\sin\varphi_2,\\
A_{33}=&0,\\
A_{34}=&H_\mathrm{ext}\cos(\varphi_2-\varphi_\mathrm{ext})-H_\mathrm{E}\cos(\varphi_1-\varphi_2)+2H_\mathrm{uni}\cos(2(\varphi_2-\varphi_\mathrm{in}))+H_\mathrm{D}\sin^2\varphi_2,\\
A_{41}=&-H_\mathrm{E}+H_\mathrm{D},\\
A_{42}=&0,\\
A_{43}=&- M_\mathrm{S}-H_\mathrm{ext}\cos(\varphi_2-\varphi_\mathrm{ext})+H_\mathrm{E}\cos(\varphi_1-\varphi_2)-2H_\mathrm{uni}\cos^2(\varphi_2-\varphi_\mathrm{in})+H_\mathrm{D},\\
A_{44}=&0.
\end{align}

\twocolumngrid

\end{document}